\documentclass[10pt,conference]{IEEEtran}
\IEEEoverridecommandlockouts

\usepackage{cite}
\usepackage{amsmath,amssymb,amsfonts}
\usepackage{algorithmic}
\usepackage{graphicx}
\usepackage{placeins}
\usepackage{subfig}
\usepackage{textcomp}
\usepackage{xcolor}
\usepackage{orcidlink}
\usepackage{tikz}
\usetikzlibrary{positioning}
\usepackage{booktabs}
\usepackage{makecell}
\usepackage{tabularx}
\usepackage{multirow}
\usepackage{arydshln}
\usepackage[super]{nth}

\newcommand{\note}[1]{}
\newcommand{\lpnote}[1]{}
\newcommand{\kwnote}[1]{}
\newcommand{\aanote}[1]{}
\newcommand{\mhnote}[1]{}

\usepackage{etoolbox}
\makeatletter
\patchcmd{\@makecaption}
  {\scshape}
  {}
  {}
  {}
\makeatletter
\patchcmd{\@makecaption}
  {\\}
  {.\ }
  {}
  {}
\makeatother
\def\tablename{Table}

\def\BibTeX{{\rm B\kern-.05em{\sc i\kern-.025em b}\kern-.08em
    T\kern-.1667em\lower.7ex\hbox{E}\kern-.125emX}}
\begin{document}

\title{Quantum Boltzmann Machines using Parallel Annealing for Medical Image Classification
\thanks{© 2025 IEEE.  Personal use of this material is permitted.  Permission from IEEE must be obtained for all other uses, in any current or future media, including reprinting/republishing this material for advertising or promotional purposes, creating new collective works, for resale or redistribution to servers or lists, or reuse of any copyrighted component of this work in other works.\\
The (partial) funding of this research by the German Federal Ministry
of Research, Technology and Space through the funding program “quantum
technologies — from basic research to market” (contract number:
13N16196) is gratefully acknowledged.}
}

\author{
\IEEEauthorblockN{Daniëlle Schuman$\orcidlink{0009-0000-0069-5517}$}
\IEEEauthorblockA{\textit{LMU Munich}\\
Munich, Germany\\
danielle.schuman@ifi.lmu.de}
\and
\IEEEauthorblockN{Mark V. Seebode$\orcidlink{0009-0004-0692-915X}$}
\IEEEauthorblockA{\textit{LMU Munich} \\
Munich, Germany\\
M.Seebode@campus.lmu.de}
\and
\IEEEauthorblockN{Tobias Rohe$\orcidlink{0009-0003-3283-0586}$}
\IEEEauthorblockA{\textit{LMU Munich} \\
Munich, Germany\\
tobias.rohe@ifi.lmu.de}
\and
\IEEEauthorblockN{Maximilian Balthasar Mansky
}
\IEEEauthorblockA{\textit{LMU Munich} \\
Munich, Germany\\
maximilian-balthasar.mansky@ifi.lmu.de}
\and
\IEEEauthorblockN{Michael Schroedl-Baumann}
\IEEEauthorblockA{\textit{SAP SE} \\
Walldorf, Germany \\
michael.schroedl-baumann@sap.com}
\and
\IEEEauthorblockN{Jonas Stein$\orcidlink{0000-0001-5727-9151}$}
\IEEEauthorblockA{\textit{Aqarios GmbH}\\
Munich, Germany\\
jonas.stein@aqarios.com}
\and
\IEEEauthorblockN{Claudia Linnhoff-Popien$\orcidlink{0000-0001-6284-9286}$}
\IEEEauthorblockA{\textit{LMU Munich}\\
Munich, Germany\\
linnhoff@ifi.lmu.de}
\and
\IEEEauthorblockN{Florian Krellner$\orcidlink{0000-0001-9532-1656}$}
\IEEEauthorblockA{\textit{SAP SE} \\
Walldorf, Germany \\
florian.krellner@sap.com}
}

\maketitle

\begin{abstract}
Exploiting the fact that samples drawn from a quantum annealer inherently follow a Boltzmann-like distribution, annealing-based Quantum Boltzmann Machines (QBMs) have gained increasing popularity in the quantum research community. While they harbor great promises for quantum speed-up, their usage currently stays a costly endeavor, as large amounts of QPU time are required to train them. This limits their applicability in the NISQ era. Following the idea of Noè et al.~\cite{noe2024quantum}, who tried to alleviate this cost by incorporating parallel quantum annealing into their unsupervised training of QBMs, this paper presents an improved version of parallel quantum annealing that we employ to train QBMs in a supervised setting. Saving qubits to encode the inputs, the latter setting allows us to test our approach on medical images from the MedMNIST data set~\cite{medmnistv2}, thereby moving closer to real-world applicability of the technology. Our experiments show that QBMs using our approach already achieve reasonable results, comparable to those of similarly-sized Convolutional Neural Networks (CNNs), with markedly smaller numbers of epochs than these classical models. Our parallel annealing technique leads to a speed-up of almost 70~\% compared to regular annealing-based BM executions.
\end{abstract}


\section{Introduction}

Machine learning and more precisely deep learning based methods have proven to be effective for medical image analysis \cite{review_Deep_Learning_in_Medical_Image_Analysis, LITJENS201760, Liu2019}, and can, for example, be used to diagnose pneumonia from chest X-rays \cite{Gupta2022}.

In the quest for near-term, attainable quantum advantage in areas like this, we propose using Quantum Boltzmann Machines (QBMs) for medical image classification. QBMs were introduced in \cite{QBM} and are the quantum analogue of classical Boltzmann Machines (BMs), a type of machine learning model introduced by Ackley, Hinton, et al. in the 1980s~\cite{ackley1985learning, optimal_perceptual_inference, A_fast_learning_algorithm_for_deep_belief_nets, Deep_Boltzmann_Machines}.

Although BMs are powerful models, they have been superseded by deep neural networks mainly because BMs are notoriously difficult to train with available classical hardware \cite{ackley1985learning, fischer2012introduction}. Therefore, current quantum research concentrates on replacing classical methods for evaluating the model states during training with quantum methods, such as Quantum Annealing (QA). Several recent work \cite{Li_2020, Kurowski_Slysz_Subocz_Różycki_2021, Niazi2024, Adachi15, dixit2020training, dorband2016boltzmann, piat2018image} has successfully used QA (or comparable techniques \cite{Niazi2024, TL_QBM_med}) in BMs trained to classify images, often finding advantages such as faster computation times \cite{Kurowski_Slysz_Subocz_Różycki_2021, Adachi15}, less fluctuations in training accuracy \cite{dixit2020training, TL_QBM_med} or a smaller amount of training epochs that is needed~\cite{TL_QBM_med}. Often, however, these approaches use rather small and simple data sets such as “Bars and Stripes”~\cite{dixit2020training} or (potentially coarse-grained) MNIST images~\cite{Adachi15, dorband2016boltzmann, Kurowski_Slysz_Subocz_Różycki_2021}.

To our knowledge, QBMs using QA have so far not been used directly to classify medical images: In \cite{TL_QBM_med} a combination of a pre-trained Convolutional Neural Network (CNN) and a QBM was used to classify medical images, with the BM being trained by Simulated Annealing (SA) as a proxy for QA, while in \cite{piat2018image} a combination of an auto-encoder and Deep Belief Network are used, where only the latter is pre-trained using a QA-based QBM.

However, while QA-based QBMs seem like a promising approach for medical image classification, a lot of the pre-existing works on QA-based QBMs (e.g.~\cite{QBM, TL_QBM_med, Stein24, dixit2020training}) do not use actual QA hardware results, as they report that training QA-based QBMs takes up prohibitively large amounts of expensive Quantum Processing Unit (QPU) time~\cite{TL_QBM_med, Stein24}. This poses a problem for the near-term applicability of such QA-approaches.

Recently, Noè et al.~\cite{noe2024quantum} presented a solution to this type of problem in an unsupervised setting, where they used Parallel Quantum Annealing (PQA)~\cite{Pelofske2022}~– a technique which embeds multiple independent problem instances in a single annealing cycle – to achieve a large decrease in required runtime. Following this idea, our approach to training a QA-based QBM in a supervised fashion tries to balance accuracy and efficiency by embedding different QBM problem instances isolated from each other onto the topology of the annealer's QPU to minimize interference. The development of our approach follows the first four stages of a 5-stage structured pipeline for quantum software engineering tailored to NISQ and early post-NISQ era applications \cite{rohe2025problem}.

The remainder of this work is structured as follows: First, Sec.~\ref{sec:background} gives some information about the workings of BMs, explains how they can be executed using QA, details the concept of PQA and subsequently introduces our approach of using improved PQA in supervised QBM training. We then briefly introduce CNNs, which will be used as a classical baseline in our experiments on two of the MedMNIST data sets~\cite{medmnistv2}. Sec.~\ref{sec:experiments} then introduces our employed data sets and performance metrics, and subsequently describes our hyperparameter optimization efforts and experimental results. Finally, our approach and results will be summarized in Sec.~\ref{sec:conclusion}, which also addresses the limitations of our present paper and how these might be addressed in future work.

\section{Models and Background} \label{sec:background}

\subsection{Boltzmann Machines for supervised learning}
\begin{figure}[hbtp]
    \centering
    \resizebox{0.35\textwidth}{0.35\textwidth}{
    \begin{tikzpicture}[
        every node/.style={draw,circle,minimum size=8mm,inner sep=1pt},
        datanode/.style={draw,circle,fill=blue!60,minimum size=8mm},
        labelnode/.style={draw,circle,fill=red!60,minimum size=8mm},
        hiddennode/.style={draw=black!70,circle,fill=gray!60,minimum size=8mm},
        textnode/.style={draw=none,font=\small}
    ]

    \def\n{8} 
    \def\radius{2.5} 

    \foreach \i in {1,...,\n} {
        \pgfmathsetmacro{\angle}{360/\n * (\i-1)}
        \pgfmathsetmacro{\x}{\radius * cos(\angle)}
        \pgfmathsetmacro{\y}{\radius * sin(\angle)}
        
        \ifnum\i<2
            \node[labelnode] (N\i) at (\x,\y) {};
        \else 
            \ifnum\i<5
                \node[datanode] (N\i) at (\x,\y) {};
            \else
                \node[hiddennode] (N\i) at (\x,\y) {};
            \fi
        \fi
    }

    \foreach \i in {1,...,\n} {
        \foreach \j in {1,...,\n} {
            \ifnum\i<\j
                \draw[gray] (N\i) -- (N\j);
            \fi
        }
    }

    \node[textnode, above=1.5cm of N1] {Visible Units};
    \node[textnode, blue!60, above=4.4cm of N7] {Input Units};
    \node[textnode, red!60, right=0.05cm of N1] {Label Unit};
    \node[textnode, black!70, below=1.4cm of N5] {Hidden Units};

    \end{tikzpicture}
    }
    \caption{Structure of a fully connected Boltzmann Machine. The network consists of visible units (blue and red) and hidden units (gray), with symmetric connections between all pairs of units. No distinction is made between layers; every node is connected to every other node. In supervised learning scenarios, some visible units (here in blue) can be used to represent the input data $d$, while others (in red, can also be multiple) can be used to represent the label $l$.}

    \label{fig:bm_structure}
\end{figure}
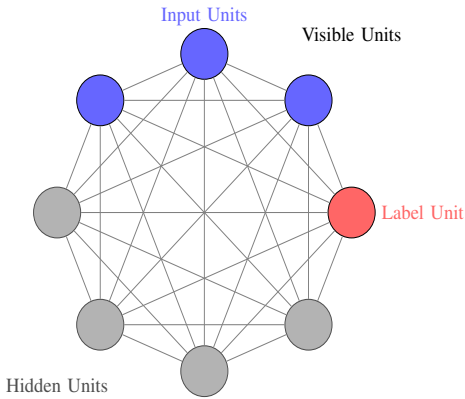

A Boltzmann Machine (BM) \cite{optimal_perceptual_inference, ackley1985learning,  A_fast_learning_algorithm_for_deep_belief_nets, Deep_Boltzmann_Machines} is an undirected stochastic neural network composed of $n$ neurons, which can be grouped into $n_v$ \textit{visible units} $v$ and (optionally) $n_h$ \textit{hidden units} $h$. They can take the values $0$ or $1$ with a certain probability governed by a quadratic energy function~\cite{ackley1985learning}. Compare Fig.~\ref{fig:bm_structure} for a visualization of such a network. 

The visible units are used to embed the input data points into the BM~\cite{QBM}. In our application, these data points consist of an input image $d$ and the corresponding label $l$. In supervised learning, a part of the visible units (blue in Fig.~\ref{fig:bm_structure}), which we will refer to in the following as \textit{input units}, will be used to encode the input data $d$ by a string of values $v_d$ of length $n_d$. We will refer to the (multi-) set of all inputs from the data set encoded this way as $D_{in}$. The remaining $n_l$ visible units (red in Fig.~\ref{fig:bm_structure}), which we will refer to as \textit{label units}, will be used to represent the label $l$ using an encoding $v_l \in \{0,1\}^{n_l}$. The goal of training the BM is that, given only a particular input $d$ (without the corresponding label $l$) at inference time, the label units will assume values $v_l$ matching the possible labels $l$ of this input with a probability closely corresponding to the frequency of the co-occurrence of the labels $l$ in question with this particular $d$ in the training data set~\cite{QBM}. This means that by drawing a few samples from the BM, the label $v_l$ most frequently assumed by the label units can be used to classify the input $d$. (Ideally, if the classification of the input can be performed unambiguously after training, the frequency of this most probable label occurring will be very close to 100\%.)

The probability of sampling a unit configuration $s \in \{0,1\}^n$ from the BM is governed by the Boltzmann distribution
\begin{equation}\label{eq:BM}
P_{b,W}(s) = \frac{1}{Z} \, \exp\left( -\frac{ E_{b,W}(s)}{T} \right)
\end{equation}
with the normalization
\begin{equation}
Z = \sum_{s \in \{0,1\}^n} \exp\left( - \frac{ E_{b,W}(s)}{T} \right) 
\end{equation}
and energy function 
\begin{equation}\label{eq:energyclassic}
 E_{b,W}(s ) = \sum_{i = 1}^n b_is_i - \sum_{\substack{i,j=1 \\ i<j}}^nW_{ij}s_is_j.
\end{equation}

where the parameters $W_{ij}$ and $b_i$ are the BM's weights and biases and $T$ is the so-called effective temperature, which can be used as a hyperparameter governing the Boltzmann distribution's shape relative to the energy function's values~\cite{ackley1985learning}: At lower effective temperatures, small differences in the energy values of states lead to more drastic differences in their probability to be sampled than at higher effective temperatures~\cite{ackley1985learning}.

The parameters of the BM are adjusted during training in a way that maximizes the likelihood of the (encoded) training data $\left( v_d, v_l \right) \in D$ (where $D$ is the encoded data set) to occur when sampling from the BM~\cite{QBM}. More precisely, following Amin et al.~\cite{QBM}, we are using a \textit{discriminative} training procedure that maximizes the likelihood of $v_l$ to be assumed by the label units when the input units are fixed, or \textit{clamped}, to $v_d$. This can be achieved by using the (average) negative log-likelihood of the labels $v_l$ as a loss function to be minimized:
\begin{equation}
    b,W \mapsto \mathcal{L}(b,W) = -\sum_{(v_d, v_l) \in D} \log P_{b,W}(v_l|v_d)
\end{equation}
Here,
\begin{equation}
P_{b,W}(v_l|v_d) = \frac{1}{Z_{v_d}} \sum_{h \in \{0,1\}^{n_h}} \exp\left( -\frac{ E_{b,W}(s)}{T} \right)
\end{equation}
with
\begin{equation}
Z_{v_d} = \sum_{(v_l, h) \in \{0,1\}^{(n_l + n_h)}} \exp\left( - \frac{ E_{b,W}(s)}{T} \right) 
\end{equation}
is the probability of sampling the label units $v_l\in \{0,1\}^{n_l}$ when $v_d$ is clamped to a specific input value~\cite{QBM}. The notation $(v_l, h) \in \{0,1\}^{(n_l + n_h)}$  denotes the concatenation of vectors $v_l$ and $h$ which has the total length $n_l + n_h$, while $s = (v_d, v_l, h)$ represents the concatenation of the vectors of all units as one vector of length $n$. Since, in this style of learning, $v_d$ is always clamped to an input value $d$ of a specific data point, the energy function $E_{b,W}(s)$ can be written as: 
\begin{equation}\label{eq:discriminative_engergy}
 E_{b,W,v_d}(s\setminus v_d) = \sum_{i = n_d + 1}^n b_i^d s_i \quad - \sum_{\substack{i,j=n_d + 1 \\ i<j}}^n W_{ij} s_i s_j
\end{equation}
where
\begin{equation}\label{eq:input_units_as_bias}
b_i^d = b_i + \sum_{k = 1}^{n_d} W_{ik} {v_d}_k
\end{equation}
acts as a bias on the on the remaining units $h$ and $v_l$~\cite{QBM}. (Here, the notation $s\setminus v_d = (v_l, h)$ is used to enhance readability.)

The standard technique to minimize $\mathcal{L}$ is via gradient descent methods. The gradient is given by~\cite{QBM}
\begin{equation}
    \partial_{b,W} \mathcal{L} = \sum_{v \in D} \langle \partial_{b,W} E_{b,W} \rangle_v \quad - \sum_{v_d \in D_{in}} \langle  \partial_{b,W} E_{b,W} \rangle_{v_d}
\end{equation}
with the Boltzmann averages
\begin{multline}
    \langle \partial_{b,W} E_{b,W} \rangle_v \\ =  \frac{1}{Z_v T}  \sum_{h \in \{0,1\}^{n_{\text{h}}}}\partial_{b,W}E_{b,W}(s) \exp{\left(-\frac{E_{b,W}(s)}{T}\right)},
\end{multline}
with normalization
\begin{equation}
    Z_v = \sum_{h \in \{0,1\}^{n_{\text{h}}}}\exp{\left(\frac{-E_{b,W}(s)}{T}\right)},
\end{equation}
and 
\begin{multline}
\langle \partial_{b,W} E_{b,W} \rangle_{v_d} \\
 = \frac{1}{Z_{v_d} T} \sum_{(v_l, h) \in \{0,1\}^{(n_l + n_h)}} \partial_{b,W}E_{b,W}\exp{\left(\frac{-E_{b,W}(s)}{T}\right)}.
\end{multline}

Thus, the gradient steps used in training the BM are given by 
\begin{equation}\label{eq:bias}
    \delta b_i = \eta (\langle s_i \rangle_v - \langle s_i \rangle_{v_d}),
\end{equation}
\begin{equation}\label{eq:weights_non_input}
    \delta W_{ij} = \eta (\langle s_i s_j \rangle_v - \langle s_i s_j \rangle_{v_d}),
\end{equation}
\begin{equation}\label{eq:weights_input}
    \delta W_{ik} = \eta (\langle s_i {v_d}_k \rangle_v - \langle s_i {v_d}_k \rangle_{v_d}),
\end{equation}
where $\eta$ is a learning rate and $s_i$ and $s_j$ are the values of hidden or label units~\cite{QBM}. In practice, these values can be determined by sampling them repeatedly from the BM in a state of equilibrium, a certain number of times in the \textit{positive} (or clamped) and an equal number of times in the \textit{negative} (or unclamped) phase~\cite{QBM}. The difference between the positive and negative phases is that in the former, in which the sample values to calculate the first terms in the Equations \ref{eq:bias} – \ref{eq:weights_input} will be determined, \textit{all} visible units will be clamped to the values $v_d$ and $v_l$ from the encoded training data point, while in the latter, used to calculate the second terms in the respective equations, \textit{only} the input units will be clamped to $v_d$~\cite{QBM}. In both cases, the pair-wise products of the sample values will then be formed to calculate the weight's gradient steps, and subsequently, the values of $\langle ... \rangle_v$ respectively $\langle ... \rangle_{v_d}$ can be calculated by averaging over the samples respectively their products~\cite{QBM, ackley1985learning}. Doing this repeatedly for all data points in the training data set will eventually cause the probability distribution of the BM (Eq.~\ref{eq:BM}) to mirror the conditional distribution of the labels (conditioned on the inputs $d$) in the training data set, which enables the BM to correctly classify the inputs $d$~\cite{QBM}.

\subsection{Using Quantum Annealing for Boltzmann sampling} \label{sec:qa-for-bm}

Reaching an equilibrium state of the BM's network which can be used to sample from can be a computationally expensive endeavor, however~\cite{fischer2012introduction, ackley1985learning}. At least using classical computers, sampling from an arbitrarily connected network requires repeatedly updating each unit according to its stochastic update rule, which involves calculating the probability of its value $s_i$ to become 1 given by
\begin{equation}
    p_i = \frac{1}{1 + \exp{(- \Delta E_{b,W}^i / T)}},
\end{equation}
where
\begin{equation}
    \Delta E_{b,W}^i = \sum_{m=1}^{n_m} W_{im} s_m + b_i,
\end{equation}
with $s_m$ being the $n_m$ neighboring units that are directly connected to the units $s_i$ in the network~\cite{ackley1985learning}. This has to be done for every unit in BM, until none of the states of the units in the network change anymore~\cite{ackley1985learning}. Furthermore, as this process has to be performed multiple times (depending on the number of samples used for averaging) for each data point, this classical training method for arbitrarily connected BMs is often considered intractable ~\cite{fischer2012introduction, ackley1985learning}.

This time consuming process of obtaining samples can, however, be avoided by using Quantum Annealing (QA) to draw samples from the BM~\cite{QBM, TL_QBM_med, Stein24}. 

D-Wave quantum annealers implement a time-dependent Hamiltonian that acts on a system of $N$ \textit{logical qubits}, interpolating between a driver Hamiltonian and a target problem Hamiltonian~\cite{rajak2023quantum}:
    \begin{equation}
    H(t) = A(t) H_D + B(t) H_P
    \end{equation}
    where $A(t)$ and $B(t)$ are scheduling functions satisfying
    \begin{equation*}
    \begin{aligned}
        A(t_i) \gg B(t_i) \text{ and } A(t_f) \ll B(t_f) \\
    \end{aligned}
\end{equation*}
at the initial ($t_i$) and final ($t_f$) times of the annealing schedule.
If the process of changing the time-dependent Hamiltonian by progressing along this annealing schedule is done adiabatically, i.e. sufficiently slow, and if the initial state at $t_i$ is the ground state of $H_D$, the resulting state at time $t_f$ should be the ground state of $H_P$\cite{rajak2023quantum}, which is to represent the most likely or best solution of a given problem, which one is looking for. Consequently, the ground state of $H_D$ should be easy to prepare \cite{rajak2023quantum}. Therefore, D-Wave's machines employ the transverse field as the driver Hamiltonian\cite{rajak2023quantum, DwaveQADocu, QBM}:
\begin{equation}
    H_D = - \sum_{i=1}^N \sigma_i^{x}
\end{equation}
with
\begin{equation}
    \sigma^{x}_i \equiv
    \underbrace{I \otimes \cdots \otimes I}_{i-1} 
    \otimes \, \sigma_x 
    \otimes \underbrace{I \otimes \cdots \otimes I}_{N - i}
\end{equation}
representing a state where the Pauli-X operator
$\sigma_x = \begin{pmatrix}
    0 & 1 \\
    1 & 0
    \end{pmatrix}$ 
acts on the $i$th logical qubit of the system
and
$
    I = \begin{pmatrix}
    1 & 0 \\
    0 & 1
\end{pmatrix}.
$
Here, $\otimes$ refers to the tensor product. 

Furthermore, the target problem Hamiltonian of interest $H_P$ takes the form of an \textit{Ising Hamiltonian} in current QA-processors~\cite{venegas2018cross}: 
\begin{equation}
    H_P = - \sum_{i=1}^N h_i\sigma^{z}_i - \sum_{\substack{i,j=1 \\ i<j}}^N J_{i,j}\sigma^{z}_i\sigma^{z}_j,
\end{equation}
with 
\begin{equation}
    \sigma^{z}_i \equiv
    \underbrace{I \otimes \cdots \otimes I}_{i-1} 
    \otimes \, \sigma_z 
    \otimes \underbrace{I \otimes \cdots \otimes I}_{N - i}
    \quad
\end{equation}
representing a state where the Pauli-Z operator
\begin{equation}
\sigma_z = \begin{pmatrix}
1 & 0 \\
0 & -1
\end{pmatrix}
\end{equation}
acts on the $i$-th logical qubit of the system and where $h_i$ and $J_{i,j}$ denote the logical qubits' biases and coupling strengths~\cite{QBM, venegas2018cross}. Ising models – just like their equivalent binary formulations, the Quadratic Unconstrained Binary Optimization (QUBO) models – are a type of energy-based model that makes it relatively easy to encode problems of which optimal, or at least good, solutions are of interest, given that these solutions can be represented by a minimum, or at least low, energy configuration of the model~\cite{venegas2018cross, QBM}. In our case, by choosing $h_i = b_i$ and $J_{i,j} = W_{i,j}$, the Hamiltonian $H_P$ can be tailored to resemble the energy function of a BM~\cite{QBM, korenkevych2016benchmarking}: Here, the logical qubits of the quantum system – which probabilistically assume values $1$ or $0$ upon measurement at the end of the annealing process – can take up the role of the hidden and label units of the network. Meanwhile, the input units – acting as biases to the rest of the network according to Eq.~\ref{eq:input_units_as_bias} – do not need to be represented by qubits and can, in theory, possess arbitrary values, including non-binary ones~\cite{QBM}.
When samples from a quantum system – i.e. strings of qubit values representing the system's configuration upon measurement – are acquired using a physical quantum annealer, the hardware interacts with the environment, resulting in a sample distribution that follows an approximate Boltzmann distribution (instead of always returning the ground state of $H_P$)~\cite{Stein24}. Thus, the QA process can be used in BM training (and inference) to reach an equilibrium from which a sample can be drawn in one step~\cite{QBM, TL_QBM_med, Stein24}. Unless explicitly stated otherwise, BMs trained and executed in this way is what we will be referring to as \textit{Quantum Boltzmann Machines (QBMs)} in the remainder of this paper. In related works on image classification tasks, such as one that utilized a Restricted QBM to pre-train a Deep Belief Network for MNIST data classification~\cite{Adachi15}, and another where classification was achieved solely with a Restricted QBM~\cite{Kurowski_Slysz_Subocz_Różycki_2021}, using Quantum Annealing for training offered benefits in computation time.

However, device-specific properties suggest that sampling from a quantum annealer's distribution will be done with an instance-dependent effective temperature, which differs from the actual hardware temperature~\cite{QBM, benedetti2016estimation}. This effective temperature can be estimated and then incorporated by rescaling the weights and biases with the inverse of said temperature~\cite{QBM, benedetti2016estimation}. Nonetheless, previous work has shown that quantum BMs trained with raw QA-samples, without any temperature correction, still arrive at probability distributions to be sampled from that sufficiently mimic the training data set's distribution to be usable for machine learning tasks, even though they might deviate from the classical Boltzmann distribution~\cite{korenkevych2016benchmarking}. Hence, this work will not incorporate a tuning of the effective temperature and instead rely on the benefit of raw QA-sampling as has been demonstrated in \cite{korenkevych2016benchmarking}.

\subsection{Employing Parallel Quantum Annealing}

When using QA hardware for BM training and inference, the logical qubits of the Hamiltonian $H_P$ must be mapped to physical qubits on the QA-hardware that have suitable connectivity to represent the connectivity of units in the BM~\cite{QBM, venegas2018cross}. This process is known as \textit{minor embedding} and is typically performed using D-Wave's \textit{minorminer API}~\cite{Minorminer, venegas2018cross}.

All experiments in this study are conducted on the D-Wave Advantage System, which contains about 5000 physical qubits and uses the so-called \textit{Pegasus topology}~\cite{quantum_hardware}. This topology supports up to 15 connections per qubit, offering improved flexibility over earlier architectures~\cite{quantum_hardware}. However, because a fully connected BM exceeds the connectivity of a single qubit, it is necessary to represent each logical unit by a chain of multiple physical qubits~\cite{venegas2018cross}. Depending on the length of the chain, this has proven to add additional noise to the system and should be avoided where ever possible~\cite{venegas2018cross}.

Minor embedding is generally applied to a single problem instance, where samples are then acquired sequentially. However, Pelofske et al.~\cite{Pelofske2022} have shown that multiple disjoint problem instances can be embedded in parallel. This approach, which we call \textit{Parallel Quantum Annealing (PQA)}, enables the annealer to generate samples from several instances within a single annealing cycle~\cite{Pelofske2022}. Their study reported significant reductions in quantum processing time, but also noted a drop in sample quality~\cite{Pelofske2022}. To address this, the authors suggested increasing the spatial distance between embeddings to reduce leakage~\cite{Pelofske2022}. Since their method relied on automatic embedding using minorminer, this separation could not be guaranteed~\cite{Pelofske2022}.

A very recent approach by Noè et al.~\cite{noe2024quantum} first also used PQA in the context of QBMs. A QBM with 16 visible and 16 hidden units was trained with the D-Wave Advantage4.1 system to reconstruct images from the 4x4-pixel “Bars and Stripes” data set in an unsupervised manner, utilizing PQA~\cite{noe2024quantum, Pelofske2022}. For the clamped phase, the researchers embedded 4 instances of the QBM into the QA hardware graph simultaneously, each instance having its visible units clamped to another data point~\cite{noe2024quantum}. For the unclamped phase, the QBM was embedded 26 times simultaneously (without any units being clamped)~\cite{noe2024quantum}. With this strategy, the quantum annealing-based training method showed an 8.6-fold improvement in wall clock time over a parallelized classical Gibbs sampling method~\cite{noe2024quantum}. In addition, their evidence shows that QBMs' sampling time scales almost linearly with the size of the model, showing potential for bigger problem instances~\cite{noe2024quantum}. However, the actual generative performance of their QBM ended up being worse than the classical approach~\cite{noe2024quantum}. Among other reasons, the authors explain these results with unsuitable temperature scaling and current hardware limitations~\cite{noe2024quantum}.

\subsection{Our Contribution: Improved Parallel Quantum Annealing for supervised learning using QBMs} \label{sec:methods}

From Noè et al.'s~\cite{noe2024quantum} description of their work, it does not become clear whether they adhered to Pelofske et al.'s~\cite{Pelofske2022} above-mentioned suggestion to actively increase the spatial distance between different problem instances – in this case different instances of the BM. Thus, it cannot be ruled out that the problems with sample quality they faced due to hardware limitations where at least partially due to unintended interactions between qubits of different problem instances that had been placed too close together on the hardware graph~\cite{Pelofske2022}.

The approach presented in this work – which is, to the best of the authors' knowledge, the \textit{first work using PQA in supervised QBM training} – avoids that limitation by \textit{explicitly controlling the placement of embeddings}:

First, to create a QBM instance to embed, we incorporate an input data point into our model. In our experiments, we do this by assigning one input unit to each of the 784 pixel values of an input from the MedMNIST~\cite{medmnistv2} data set that is used. It is important to note that the input units do not necessitate specific hardware resources since they are permanently fixed to the data, effectively serving as biases for the other nodes as described in Eq.~\ref{eq:input_units_as_bias}. As we focus on binary classification, we then embed the class label by adding one label unit to the visible layer – encoding our two classes as 0 and 1. Furthermore, for the number of hidden units, we use a parameter that takes values between 1 and 20. We purposefully chose this relatively low upper limit for the number of hidden units, since this choice avoids exposing the model to unnecessary noise which otherwise might have been caused by excessively long chains of physical qubits that have to be build when embedding the model into hardware. At the end of these steps, we have created a QBM model instance with a maximum of 21 fully connected qubits to embed, as only the hidden units and the label units will need to be embedded onto the QA-hardware. This can be done using a QUBO model of the instance, which essentially takes the same shape as the QBM's energy function displayed in Eq.~\ref{eq:discriminative_engergy} and can easily be mapped to a Hamiltonian $H_P$ using the D-Wave API~\cite{DwaveQceanSDK}.

\begin{figure*}[hbtp]  
    \centering
    \subfloat[Regions and buffer zones]{
    \includegraphics[width=0.35\textwidth]{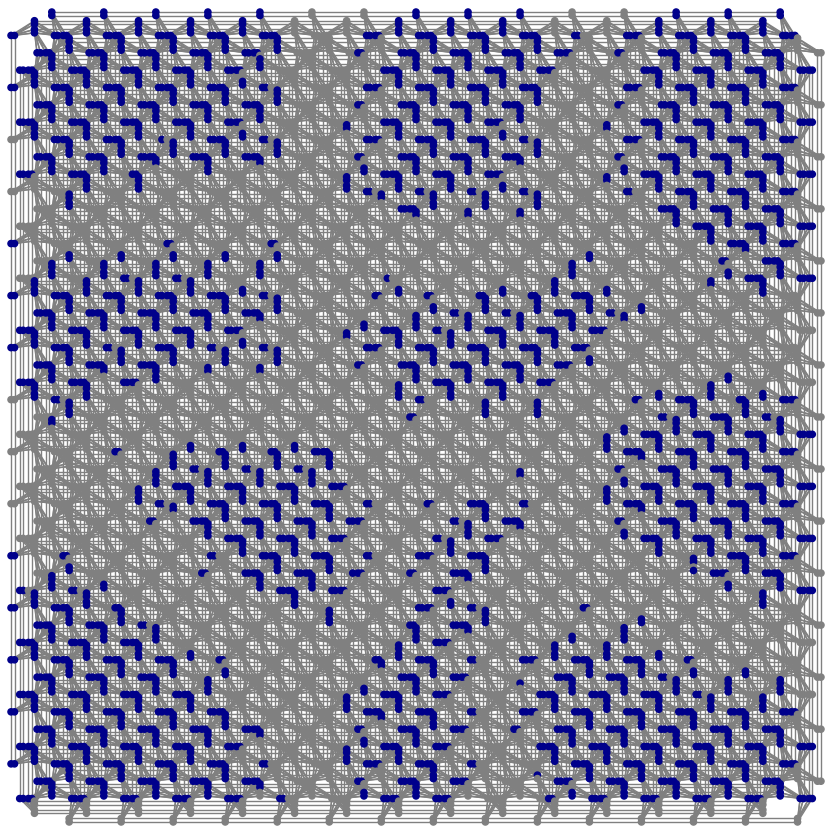}
    \label{fig:subgrahpsFig}
    }
    \hspace{0.07\textwidth}
    \subfloat[Resulting embedding]{
    \includegraphics[width=0.47\textwidth]{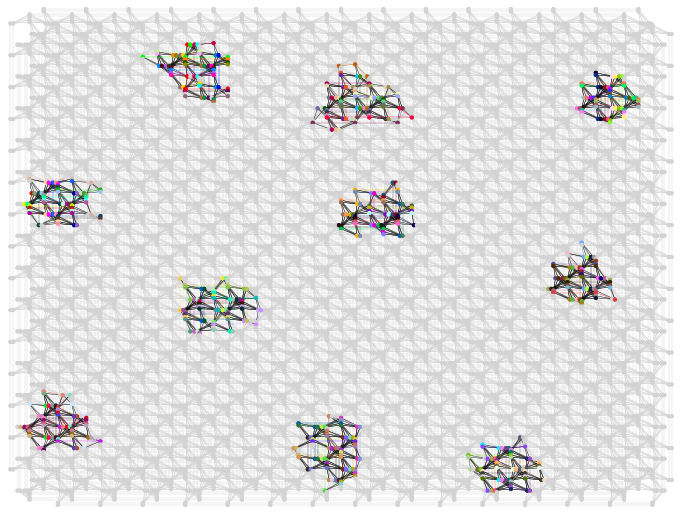}
    \label{fig:multiembedding}
    }
    \caption{Visualization of the parallel embedding approach on the Pegasus topology. (a) The topology is partitioned into 10 subgraphs. Blue nodes indicate the regions used for embedding, while gray nodes represent buffer zones that enforce a minimum distance between embeddings. (b) An example embedding of a 20 hidden unit QBM (QA) in the clamped phase, demonstrating how the parallel embedding strategy utilizes the partitioned regions.}
    \label{fig:parallel_embedding_approach}
\end{figure*}

Given our QBM instance as a QUBO, we then divide the Pegasus topology of the D-Wave Advantage's hardware graph into ten subgraphs using the \textit{Pymetis} Python library ~\cite{pymetis2022}. This fixed number of subgraphs ensures sufficient possible spacing between the embeddings, given the maximum of 21 fully connected qubits we need to embed. 
To further isolate each of the ten subgraphs, we introduce a buffer zone between them: Nodes that are connected to other subgraphs are removed to reduce the likelihood of interference between embeddings. The final topology, including both the subgraphs (blue) used for finding embeddings and buffer zones (gray), ensuring a minimum distance between the embeddings, is shown in Fig.~\ref{fig:subgrahpsFig}. Each subgraph is then used to embed the same QUBO model of the QBM instance using minorminer. Fig.~\ref{fig:multiembedding} shows an example of this multi-embedding, which includes a clamped QBM instance with 20 hidden units. (Here, the label units do not need to be embedded, given that in the clamped phase, they act as biases in the same way the input units do.) Using this type of multi-embedding, we can subsequently draw ten samples at once using one execution of the annealing cycle. While this strategy ensures a minimum distance between embeddings, thus preserving sample quality, it also results in a significant number of qubits being allocated to buffer zones. This reduces the overall embedding capacity of a given QA hardware graph.

In order to successfully evaluate the performance of our supervised QBMs that employ PQA and see how the approach scales when employing different model sizes, we perform experiments on two of the MedMNIST data sets~\cite{medmnistv2}, see Sec.~\ref{sec:experiments}. Depending on the number of hidden units used, the configurations explored here contain between 1571 and 16695 trainable parameters: The minimum of 1571 parameters includes $784 \cdot 2 = 1568$ weights (acting as biases according to Eq. ~\ref{eq:discriminative_engergy}) from the input units to the rest of the network, one weight between the single hidden unit and label unit and one bias for each of those units. Similarly, the maximum of 16695 parameters includes $784 \cdot 21 = 16464$ input weights, $(20 \cdot 19)/2 = 190$ weights between the hidden units, $20$ weights between the hidden units and the single label unit, $20$ hidden biases and one label unit bias. In all cases, all parameters are initialized with random values drawn uniformly from the interval $[-1, 1]$. To also compare the performance of these differently sized QBM models not just to each other, but also to that of some classical models of a similar size that are commonly used for the task at hand, we evaluated them against Convolutional Neural Networks (CNNs) using a similar number of parameters~\cite{LeCun1989GeneralizationAN, alex_net}. These CNNs are described in the following section.

\subsection{Convolutional Neural Networks} \label{sec:CNNs}

Deep Convolutional Neural Networks (CNNs) \cite{LeCun1989GeneralizationAN} became the state of the art for image classification with the introduction of AlexNet \cite{alex_net}. CNNs are made of two types of layers. First, convolutional layers that apply filters to input data to extract feature maps, and pooling layers that downsample these feature maps to reduce their dimensionality~\cite{lecun2015deep}. Second, additional fully connected layers to make predictions based on the extracted features~\cite{alex_net}. 

There is considerable empirical evidence that CNNs with more parameters tend to be more performant on many tasks. With $60$ million parameters, AlexNet \cite{alex_net} was the first deep CNN that showed unprecedented success on large-scale and complex tasks, winning the 2012 ImageNet competition \cite{SimonyanZ14a_very_deep}. In \cite{SimonyanZ14a_very_deep} VGG networks were introduced. The publication demonstrated that increasing network depth significantly improved performance on the ImageNet challenge~\cite{SimonyanZ14a_very_deep}, providing evidence that deeper networks, and hence networks with more parameters, can learn more powerful representations. The VGG16 model has approximately $138$ million parameters~\cite{SimonyanZ14a_very_deep}. In \cite{resnet}, residual networks were introduced. This work demonstrated that deeper ResNet models perform better than shallower ones~\cite{resnet}. The smaller ResNet-18 model has (about) $11.7$ millions parameters~\cite{resnet_pytorch}; a number not being able to be matched with QBMs, due to the lack of sufficient quantum hardware. 

To fairly compare QBMs with CNNs, we propose a CNN architecture with one convolutional layer and two sequential layers. In this setup, we are able to vary the number of parameters in a range similar to that of the QBM. The kernel size of the convolutional layer is $3$ or $5$ and always of dimension $1$. The number of neurons of the first sequential layer is in $\{4,8,16,24\}$ and for the second layer, it is in $\{2,4,8,16\}$, with the number of neurons of the second sequential layer never being higher than that of the first one.

After the convolutional layer and the first sequential layer we apply the 
\begin{equation}
    \text{ReLU}(x)=\max(0,x)
\end{equation}
activation function. After the last sequential layer, the sigmoid activation
\begin{equation}
    \sigma(x) = \frac{1}{1+\exp{(-x)}}
\end{equation}
is applied, mapping the output to the interval $[0,1]$ and making it interpretable as a probability. 

We use the binary cross entropy loss 
\begin{equation}
    \ell(x,y) = - \left( y \log(x) + (1-y)\log(1-x) \right)
\end{equation}
for training our CNN. Let 
\begin{equation}
    \text{cnn}_{\theta}:[0,1]^{28,28} \rightarrow [0,1] 
\end{equation}
be the CNN and 
\begin{equation}
    (x_i,y_i)_{i=1,\ldots,n} \in [0,1]^{28,28}\times \{0,1\}
\end{equation}
be our encoded data set, then the training of our CNN is the following optimization problem:
\begin{equation}
    \min_{\theta} \sum_{i=1}^n\ell(\text{cnn}_{\theta}(x_i), y_i).
\end{equation}

Such optimization problems are solved with gradient-based methods. We used the Adam optimizer for training \cite{adam}.

\section{Experiments}\label{sec:experiments}

\subsection{The MedMNIST data sets and their common performance metrics}

All experiments are conducted using two datasets from the MedMNIST  \cite{medmnistv2} collection of biomedical images: Pneumonia-MNIST and BreastMNIST. Both datasets are preprocessed into grayscale images of size $28 \times 28$ pixels and provide a standardized train-validation-test split.

The PneumoniaMNIST dataset consists of 5856 pediatric chest X-ray images, categorized into two classes: pneumonia and healthy. It is derived from the substantially larger images in \cite{kermany2018identifying}, from which the images where center-cropped and subsequently resized to match the low-resolution format. The data is divided into training and validation sets in a 9:1 ratio, with the original source validation set used as the test set.

The BreastMNIST dataset includes 780 breast ultrasound images and is based on $ 500 \times 500$ pixels large images from \cite{breastmnist}. Due to the significantly lower resolution, the original three-class classification task (normal, benign, malignant) was reformulated as a binary classification problem, “normal” and “benign” being merged into a single positive class, while the “malignant” was treated as negative. The dataset is split into training, validation, and test sets using a 7:1:2 ratio.

Medical image datasets often exhibit substantial class imbalance, which can lead to misleading performance evaluations when using the standard accuracy (ACC) as a primary metric. For instance, approximately $73 \%$ of images in both PneumoniaMNIST and BreastMNIST are labeled positive, which means a constant estimator would achieve an accuracy of $73 \%$.

The AUC-Score, which we, as well as \cite{medmnistv2}, use in addition to the ACC, is derived from the ROC curve, which plots the true positive rate against the false positive rate at various thresholds. This metric can thus reflect the models' ability to distinguish between class labels: A score of $0.5$ indicates random guessing, whereas a score of 1.0 indicates perfect separation of class labels. 

\subsection{Hyperparameter optimization}

\subsubsection{QBM}

The performance of a BM is influenced by several hyperparameters, which makes an appropriate hyperparameter optimization essential. The selected hyperparameters and their corresponding search spaces are summarized in Table \ref{tab:hyperparameters_qbm}.

\begin{table*}[hbtp]
\caption{Hyperparameter Ranges for the QBM (SA) and the Best Hyperparameter for Each Dataset}
\centering
\begin{tabular}{|c|c|c|c|c|c|}
\hline
\textbf{ } & \textbf{Hidden Units} & \textbf{Epochs} & \textbf{Batch Size} & \textbf{Learning Rate} & \textbf{Sample Count} \\
\hline
Value ranges & 1--20 & 1--20 & 1--100 & 0.00001--0.6 & 10--1000 \\
\hline
PneumoniaMNIST & 10 & 20 & 73 & 0.45295 & 100 \\
BreastMNIST    & 8  & 13 & 12 & 0.43496 & 400 \\
\hline
\end{tabular}
\label{tab:hyperparameters_qbm}
\end{table*}

We initially aimed to perform 100 hyperparameter optimization runs per data set, using 10 different random seeds each for, among other things, weights initialization. The results from these runs were to be averaged across seeds, and the optimization was conducted using the Bayesian search algorithm provided by the \textit{Weights and Biases} framework \cite{wandb}. In order to have the optimization respect both the accuracy and the AUC-score, we employed a composite score defined as $0.5 \cdot \text{ACC} + 0.5 \cdot \text{AUC}$. However, conducting such extensive optimization using QBMs with QA would have exceeded our available access to D-Wave QA-hardware. Prior work \cite{TL_QBM_med, Stein24} has demonstrated that for the purpose of hyperparameter tuning, the Simulated Annealing algorithm (SA) can serve as a practical classical alternative to QA, as it also approximates a Boltzmann distribution \cite{nishimori2015comparative}. Despite employing this alternative as as a workaround, long algorithm run times and unexpected failures of our classical hardware unfortunately limited the number of optimization runs we were able to complete in time for the preparation of this paper to only 20 runs on the BreastMNIST dataset and only 18 runs on the PneumoniaMNIST dataset regardless, meaning that not much optimization of hyperparameters took place in the end.

Still, the models achieving the best validation performance were then chosen to be retrained with QA, using our PQA strategy and 3 random seeds.

\subsubsection{CNN}

As with the QBM, we also used hyperparameter optimization to find the best architecture and optimization parameters for the CNN. The selected hyperparameters and the search space are summarized in Table~\ref{tab:hyperparameters_cnn}. 

For the hyperparameter search for the CNNs we combined grid search~\cite{hyperparameter_grid_search} with random search~\cite{hyperparameter_random_search}: For every pair of choices for kernel size and the number of neurons in the two consecutive sequential layers, making sure that the second layer has no more neurons than the first, we randomly picked $200$ values for each of the other hyperparameters. Then, we picked the configuration that produced the highest combined score on the validation set – that is, the sum of the AUC-score and accuracy – during training. 

The number of epochs was chosen high enough such that the validation performance stopped improving at the end of the training. Specifically, we ran $50$ epochs for PneumoniaMNIST and $350$ epochs for BreastMNIST. This difference is due to the datasets' sizes; using more epochs for BreastMNIST ensured roughly the same number of gradient steps during training, given an equal batch size.

\begin{table*}[hbtp]
\vspace{0.4cm}
\caption{Hyperparameter Ranges for the CNN and the Best Hyperparameter for Each Dataset}
\vspace{-0.185cm}
\begin{center}
\begin{tabular}{|c|c|c|c|c|c|c|c|}
    \hline
    &\textbf{Kernel Size} & \thead{\textbf{Neurons} \\ \textbf{\nth{1} Sequential} \\ \textbf{Layer}} & \thead{\textbf{Neurons} \\ \textbf{\nth{2} Sequential} \\ \textbf{Layer}} & \textbf{Learning Rate} & $\mathbf{\beta_1}$ & $\mathbf{\beta_2}$ & \textbf{Batch Size} \\
    \hline
    Value ranges & $ 3,5$ & $ 4,8,16,24 $ & $ 2,4,8,16 $ & $[0.0005,0.05]$ & $[0.998,0.9999]$ & $[0.98,0.999999]$ & $2^2,\ldots,2^{10}$ \\
    \hline
    PneumoniaMNIST & $5$ & $16$ & $8$ & $0.00384$ & $0.98428$ & $0.99925$ & $16$\\
    BreastMNIST & $5$ & $24$ & $8$ & $0.00117$ & $0.98674$ & $0.99931$ & $8$\\
    \hline
\end{tabular}
\end{center}
\label{tab:hyperparameters_cnn}
\end{table*}

\subsection{Results}

\begin{figure*}[hbtp]
    \centering
    \subfloat[PneumoniaMNIST]{
    \label{fig:scatter_pneumonia}
    \includegraphics[width=0.49\textwidth]{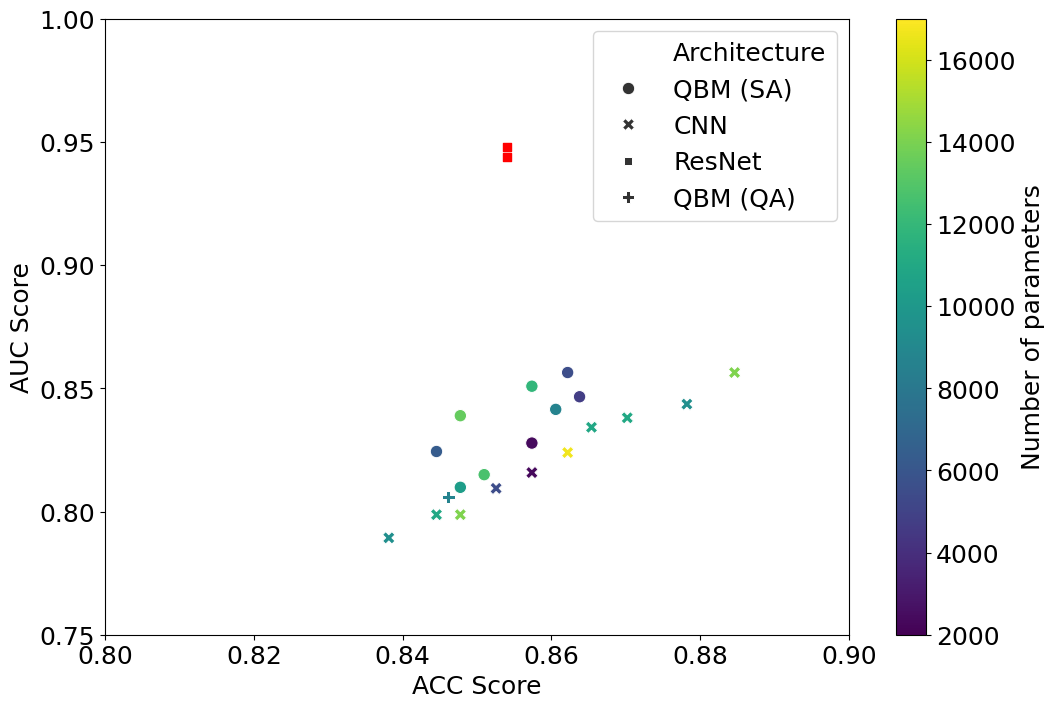}}  
    \subfloat[BreastMNIST]{
    \label{fig:scatter_breast}
    \includegraphics[width=0.49\textwidth]{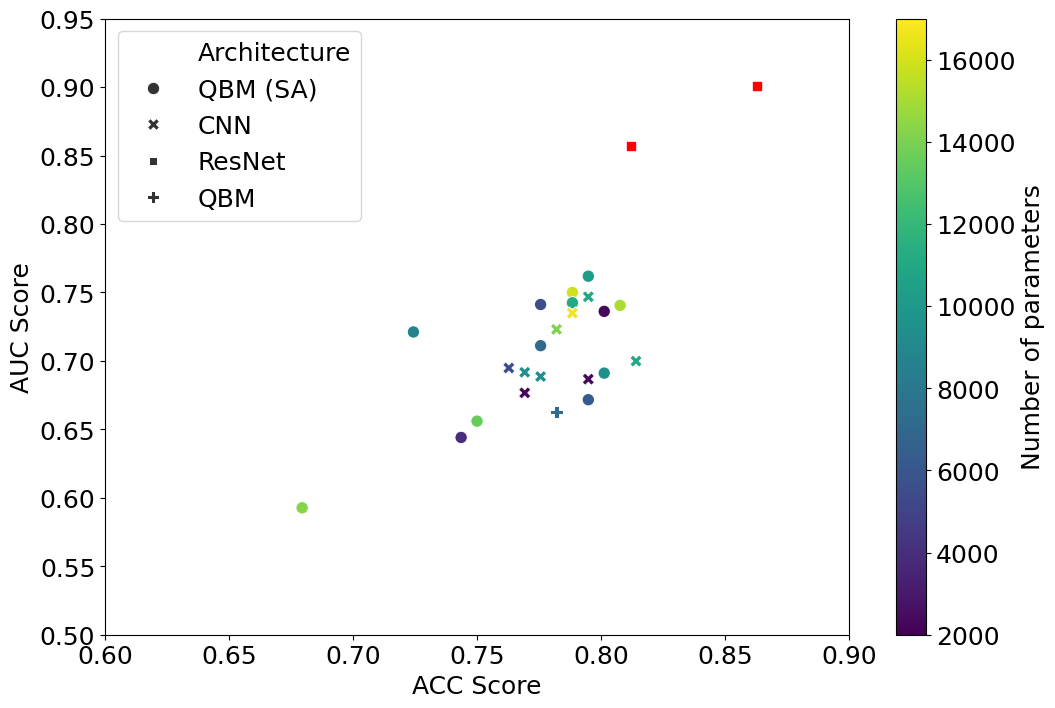}}
    \caption{The figures compare the performance of the QBMs and CNNs on the PneunomiaMNIST and the BreastMNIST data sets. In each panel, the models' test AUC and ACC-scores are presented, while a color map indicates the number of parameters. For additional context, we included the results for two variants of the ResNet CNN \protect\cite{resnet} as reported in \protect\cite{medmnistv2}. They are highlighted in red because their parameter counts exceed the colormap’s representable range: ResNet-18 has $11.69$ million parameters~\protect\cite{resnet_pytorch}, and ResNet-50 about $25.56$ million~\protect\cite{resnet_50_pytorch}.}
    \label{fig:results_mnist}
\end{figure*}

As outlined in Sec.~\ref{sec:methods}, our experimental objectives were to evaluate how our QBM-based approach compares to CNNs of equivalent size, which are traditionally used in image classification, as well as whether its performance scales with different model sizes. Some preliminary answers to both of theses questions can be found in Fig.~\ref{fig:results_mnist}: For each of the different numbers of hidden neurons explored in our respective hyperparameter optimizations, using the “QBM” with SA (shown as circles) as well as the CNN (shown as crosses), the plot displays the accuracy (ACC) and AUC-Scores of the best models found using these neuron numbers, with respect to the remaining hyperparameters mentioned in Tables~\ref{tab:hyperparameters_qbm} and \ref{tab:hyperparameters_cnn} as well as the random seed used for initialization, on the test data sets. Notice that in the context of the “QBM” trained with SA, one cannot really speak of the hyperparameters of each of these configurations having been optimized – given that only very little runs per number of hidden neurons have been performed. Still, we suspect that even the performance metrics of the models that did not or only barely undergo hyperparameter optimization allow for some preliminary insights. Furthermore, the plots also show the best overall result obtained by running the QBM on quantum hardware (shown as +), as well as the results of the millions of parameters large state-of-art CNNs ResNet-18 and ResNet-50 from literature~\cite{medmnistv2}, shown in red. For the PneumoniaMNIST data set, displayed in Fig.~\ref{fig:scatter_pneumonia}, we observe no clear trend of parameter numbers much influencing model performance for either of the models – with the exception of the (surely also in other ways optimized) ResNet approaches reaching significantly higher AUC-Scores. While all other approaches show AUC-Scores in a very similar range, we find that the CNN models can generally still outperform their QBM-based counterparts by a bit in terms of accuracy on this data set – some even improving upon ResNet in this regard. The best CNN variants also outperform their similarly sized QBM counterparts regarding the AUC-Score they reach, the overall best models matching each others performance in this regard. While having the CNNs outperform the QBMs is not surprising, given – among other things – their more extensive hyperparameter optimization, an interesting finding in this plot is that while still outperformed by most other models, the QBM trained with actual QA is fairly close in performance to some of the similary sized SA-trained models, and even outperforms some similarly sized CNNs, both in terms of ACC and AUC-Score. On the smaller BreastMNIST data set (Fig.~\ref{fig:scatter_breast}), however, results look vastly different in a lot of aspects: While here, the SA-trained QBMs still do not show a clear trend regarding a possible interconnectedness of performance and parameter number, CNNs do show a fairly clear trend regarding their improvement with size, with ResNet also clearly showing the best performance regarding both metrics this time. Additionally, the QA-based QBM is again noticeably outperformed by the majority of other the examined models, particularly the slightly larger CNNs, when evaluating AUC Score. However, it demonstrates a moderately average ACC, surpassing many of the smaller models in this regard. Quite some of SA-based QBMs do, however, outperform their similarly – or a bit larger – sized CNN counterparts here. Taken together, we do not see any clear conclusions that can be drawn from these two experiments about the general (medical) image classification performance of QBMs in comparison to CNNs just yet. We can, however, say that the number of parameters does not seem to have a huge impact on model performance for QBMs (even though we do not know with certainty that this would not change when employing more hyperparameter optimization).

\begin{figure*}[hbtp]
    \centering
    \subfloat[Average test accuracies PneumoniaMNIST]{
    \label{fig:test_acc_per_epoch_pneumonia}
    \includegraphics[width=0.49\textwidth]{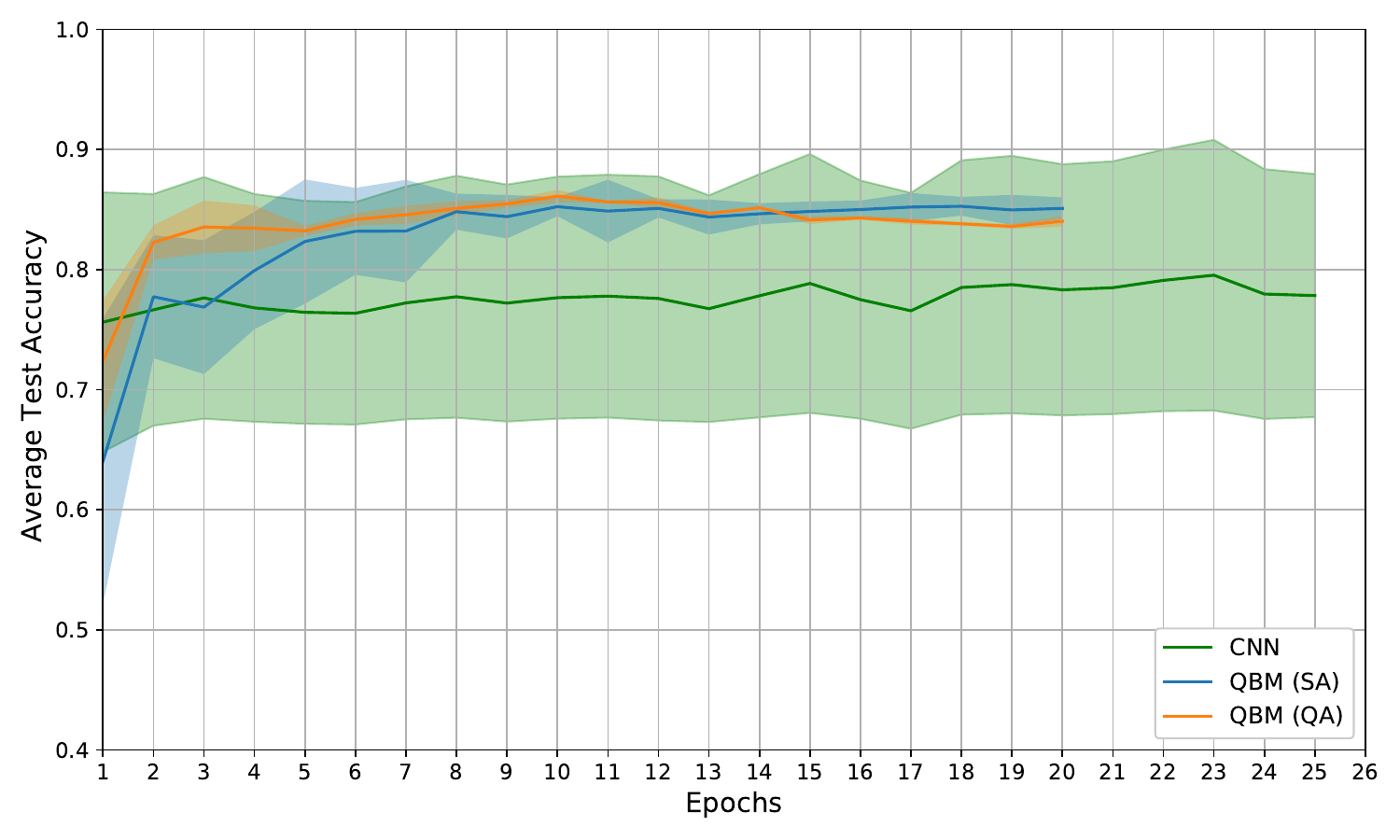}} 
    \subfloat[Average AUC-Score PneumoniaMNIST]{
    \label{fig:test_AUC_per_epoch_pneumonia}
    \includegraphics[width=0.49\textwidth]{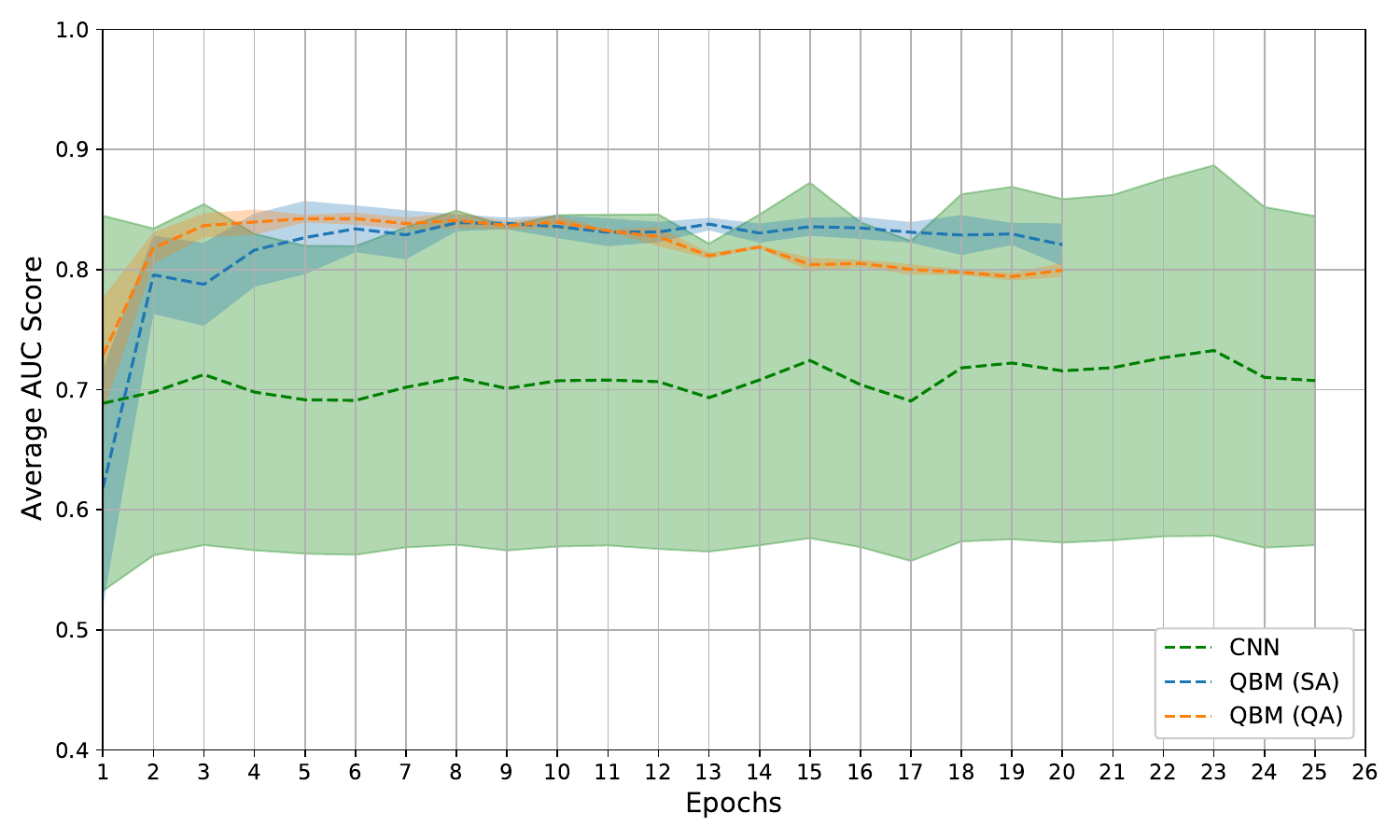}}\\
    \subfloat[Average test accuracies BreastMNIST]{
    \label{fig:test_acc_per_epoch_breast}
    \includegraphics[width=0.49\textwidth]{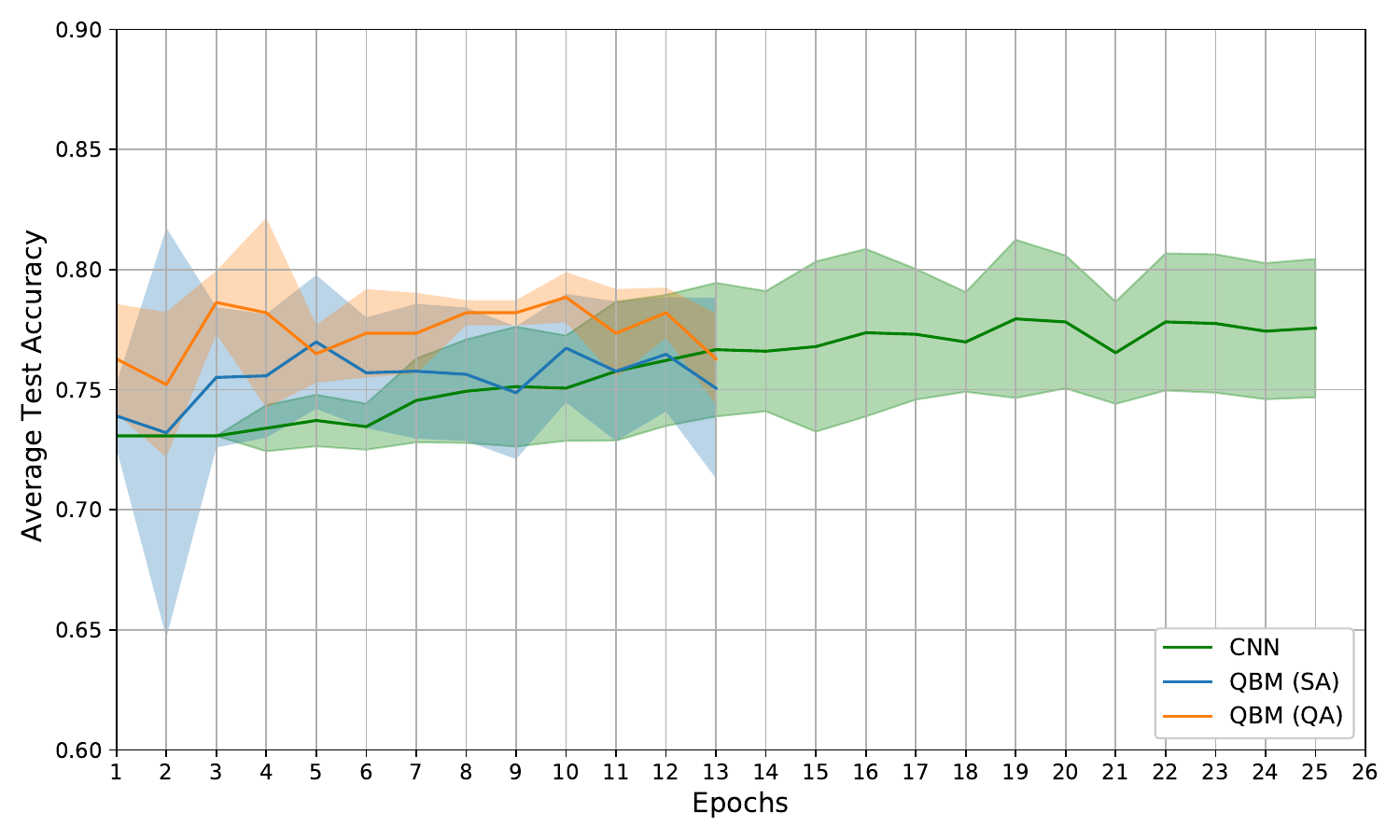}}
    \subfloat[Average AUC-Score BreastMNIST]{
    \label{fig:test_AUC_per_epoch_breast}
    \includegraphics[width=0.49\textwidth]{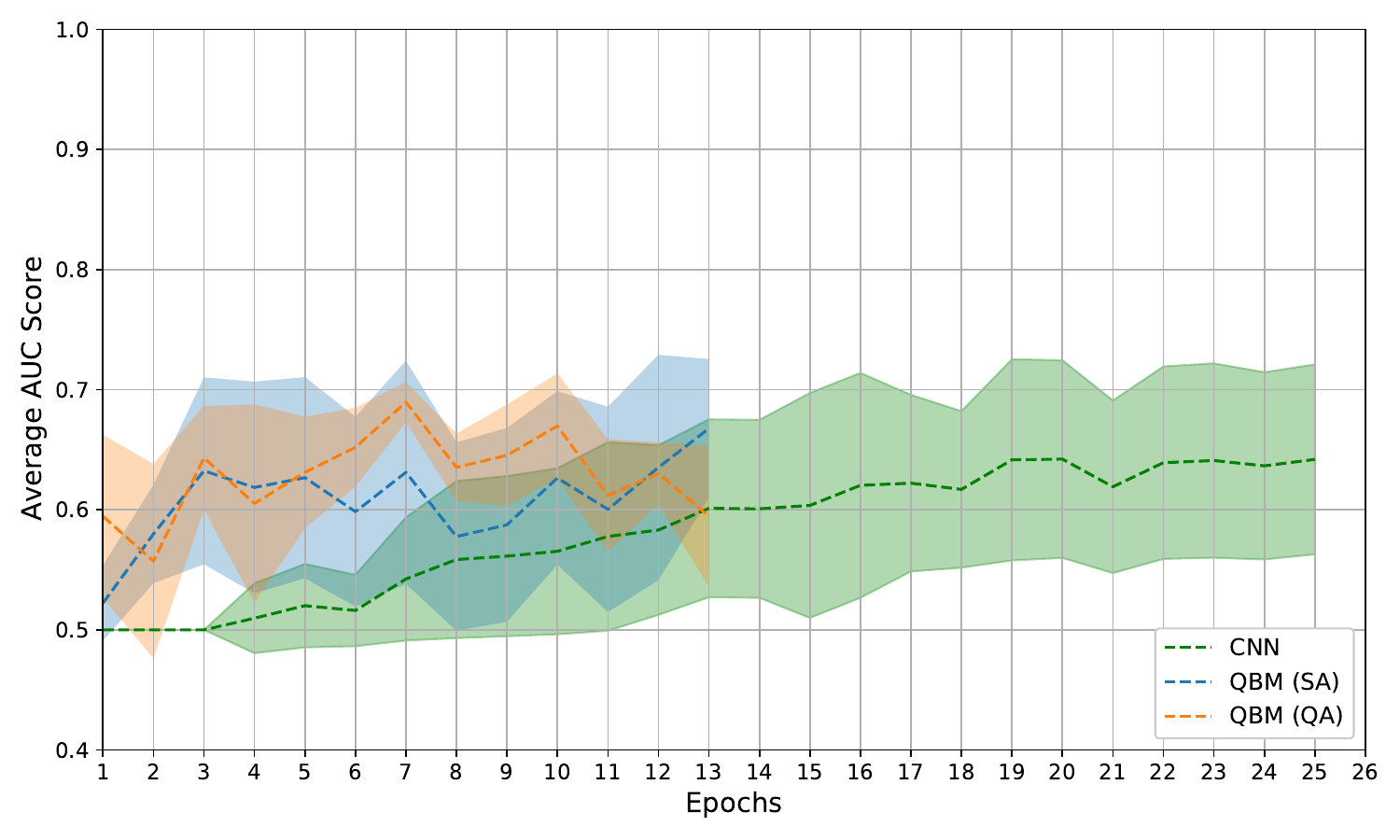}}
    \caption{Test accuracies and AUC-Scores of QBMs trained using SA and QA, as well as a classical CNN with the best identified hyperparameters settings, per epoch. The solid lines represent the average performance over 10 (QBM~(SA) \& CNN) respectively 3 (QBM (QA)) independent seeds. The transparent area around them is the standard deviation.}
    \label{fig:test_auc_per_epoch}
\end{figure*}

Thus, we want to take a closer look at the average performance of the models with the absolute best hyperparameter configurations we found so far (regarding classification performance on the validation set after the last epoch) – independent of parameter size and initialization with random seeds. The selected hyperparameters used for this are summarized in Table~\ref{tab:hyperparameters_qbm}. As one of our previous works on medical image classification using SA-based QBMs~\cite{TL_QBM_med} found signs that the usage of QBMs might not only reduce sampling time, but also the number of epochs necessary to reach good classification performance, we investigate the classification performance on the test set after each epoch of training when doing so. 

The result, showing the average test accuracies and average test AUC-Scores across different random seeds, as well as their standard deviations, can be seen in Fig.~\ref{fig:test_auc_per_epoch}. We would like to point out that, as the QBM (QA) was only run with three seeds due to machine-failure-related time constraints, we do not consider its standard deviation representative enough to draw any conclusions from that. To make at least the results for the QBM (SA) and the CNN directly comparable here, only the results of ten random seeds are plotted here for both models. Comparing the standard deviations for both these models, we notice, at least on the Pneumonia data set, that the CNNs standard deviation is a lot larger. This increased variability is due to the fact that in multiple training runs, the CNN consistently predicted the majority class, which led to the gradients vanishing — a common issue when training CNNs~\cite{vanishing_gradient}. Although these vanishing gradients affected the average performance, they did not impact the highest performance achieved in our training runs. We do not observe this effect on the BreastMNIST data set, however, where the standard deviations are fairly comparable. 

Comparing the average performance metrics on the PneumoniaMNIST data set, we notice that both QBM versions seem to clearly outperform the CNN both in terms of ACC as well as in terms of AUC-Score – at least in the first 25 epochs displayed here. Remarkably, they already reach their high classification performances within the first 8 epochs – even the first 5 ones for the QBM (QA). A similar observation can be made for the BreastMNIST data set: While here, the CNN seems to be “catching up” with the QBMs over the course of the training, at the very least with the QBM (SA) in terms of accuracy, the QBMs clearly outperform the CNN in the first 10 epochs of training in terms of both metrics – again with the QA-based version reaching its maximum performance (for the first time) earlier than the SA-based one. Within the first 7 epochs of training, the QBM (QA) outperforms the QBM (SA) in terms of both metrics, and even in the longer run it still reaches a comparable performance. This suggests that the QBM (QA) suffers very little from hardware noise in this type of application, or perhaps not at all.

Regarding absolute classification performance, the QBMs also provide decent results on the PneumoniaMNIST data set: The average test accuracy at the final epoch in case of the QBM (SA) is $85.10\%$ and the AUC-Score is $0.8208$, the second value of which almost comes close to values reached by large models like ResNet in literature~\cite{medmnistv2}, while our QBM (QA) reaches $84.03\%$ and $0.7996$, respectively. In case of the BreastMNIST dataset, the performance of all models deteriorates significantly, however. The QBM (SA) achieves a test accuracy of $75.06\%$ and an AUC-Score of $0.6677$, while the QBM (QA) reaches $76.28\%$ and an AUC-Score of $0.5946$. This decline is likely attributable to the limited size of the training set (546 images~\cite{breastmnist}), posing challenges for effective generalization.

\begin{figure}[hbtp]  
    \centering
    \includegraphics[width=0.42\textwidth]{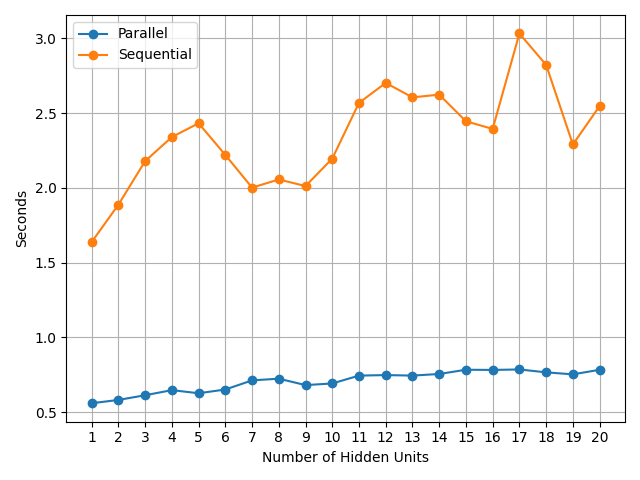}
    \caption{The plot shows the total QPU time required to perform 3 batches of 5 steps each, with every step generating 1000 samples for both the clamped and unclamped phase of QBM (QA) training. Results are presented for both standard sequential QA and our approach using PQA.}
    \label{fig:sequential_vs_parallel}
\end{figure}

In addition to evaluating the classification performance of our QBMs, we also conduct a small analysis of our PQA approach in terms of QPU time: For QBM configurations of up to 20 hidden units, we tracked the QPU time expenditure of QBM training with both regular sequential, as well as our parallel QA, in seconds for 3 mini-batches à 5 data points each, generating 1000 samples for both the clamped and unclamped training phases with each data point. The results are shown in Fig.~\ref{fig:sequential_vs_parallel}. Here, PQA unsurprisingly exhibits a much more favorable QPU time usage compared to sequential QA: Averaged over every possible configuration, our PQA approach exhibits a speed up of $69.65\%$. Furthermore, it also seems to scale more stably and favorably with increasing network sizes. 

\section{Discussion And Conclusion} \label{sec:conclusion}

In this work, we presented an approach for image classification using quantum annealing (QA)-based Quantum Boltzmann Machines (QBMs) which are efficiently trained in a supervised manner using Parallel Quantum Annealing (PQA): Our approach involves simultaneously embedding several instances of the QBM into a quantum annealer's hardware graph by embedding them into artificially separated subgraphs. The resulting distance of the instances' embeddings is thought to preserve sample quality~\cite{Pelofske2022}, while enabling the usage of QA to draw several Boltzmann samples for the QBM's training process in parallel. This way, we achieve a speed-up of $69.65\%$ compared to the usage of regular sequential QA. Although our experiments remain inconclusive regarding the QBM's ability to outperform classical CNN models of similar size in terms of classification performance, we find that they on average can reach decent classification metrics, similar to those of the CNNs, within much shorter numbers of epochs. Taking together these two aspects, we support future research into the proposed approach, as we deem it not unlikely that a future quantum speed-up could be achieved in this area.

Future work should first address the limitations of our current research: 

Our experiments should be repeated, this time involving a consistent and extensive approach to hyperparameter optimization, especially for the QBM, to ensure that the best possible performance of each type of model is actually obtained. Ideally, the optimization should already be executed using QA, as SA, albeit resulting in a similar sample distribution, does not outright simulate it and thus does not necessarily return the exact same parameters. The search also should include the optimization of $T$, which we consciously left out in this work due to the limited availability of QPU time. Furthermore, large numbers of random seeds should be used across all models to enable statistically significant comparisons of e.g. their sensitivity to “bad” weight initializations for a given data set. And moreover, the ranges of hyperparameter search should be extended, e.g. to investigate a QBM's performance using larger numbers of epochs as well. 

Secondly, it would be desirable to execute experiments on future generations of hardware, which offer more qubits and a denser connectivity, to enable the embedding of larger QBMs, in terms of hidden units, while guaranteeing similar or lower levels of noise. 

And thirdly, it would be desirable to benchmark our QBM on datasets containing larger images or multiple classes, to see how its performance scales in comparison to that of classical models like (larger) CNNs with increasing problem complexity. Since, as mentioned above, input units only function as bias terms to the remaining units of the QBM within the framework of discriminative learning, the number of units in our model that need to be represented by logical qubits, which are embedded into the quantum annealer's hardware graph, is determined solely by the sizes of the hidden and label layers. Thus, the additional input units required to represent larger input images, with more pixels, would not impact the size of the embedded QBM instances our model uses. Furthermore, when increasing the number of classes, the number of output units needed scales only linearly with the number of classes our model is supposed to be able to represent, as we need one output unit per class. This means that unless the number of classes in the data set becomes exceedingly high – which might be the case for large generic data sets like ImageNet~\cite{imagenet}, but is not expected for medical ones – increasing the number of classes does not have much negative impact on the embeddability of the QBM instances either. It remains to be investigated whether increasing the number of hidden units, in order to increase the parameter count of the QBM, would be necessary to ensure that a desirable classification performance can be reached on these more complex types of data sets. However, our preliminary results on the data sets used in this work suggest that increasing or reducing parameter count – and thus the number of hidden units required – might not have as much of a critical impact on our QBM model as it seems to usually have on some classical models like CNNs. Thus, even though it also remains to be investigated how the potential benefits we see in terms of the rather low number of epochs needed in training scale with increasing data complexity, we would expect our approach to scale rather well on more complex data sets, as the PQA should still be usable without difficulty. Whether the approach could then outperform larger CNNs on these tasks in terms of classification performance or training speed, e.g. by again needing only a smaller number of epochs, remains to be seen.

And finally, it would be interesting to compare the QBM as we use it in our work to more similar classical models, given that the CNNs we use as a classical baseline in this work, while being especially well suited to the task of image classification, process these images in a very different way that makes use of the spacial structure of the input features (compare Sec.~\ref{sec:CNNs}), while the QBM works on flattened inputs. Thus, while CNNs might be a good baseline to investigate whether the QBMs perform well, a comparison against them is not particularly well suited to investigate whether any potential advantages we see when using our model stem from using quantum computing. While of course we already present one of the most direct classical comparisons possible in this work in form of the QBM (SA) model, it might be interesting in this context to also compare our QBM to more common comparable classical models such as (Classification) Restricted Boltzmann Machines~\cite{larochelle2008classification, larochelle2012learning}, classical general Boltzmann Machines trained using typical Boltzmann learning~\cite{ackley1985learning}~\footnote{While general Boltzmann Machines are sometimes also said to be trained with “Simulated Annealing” (compare e.g.~\cite{hinton1986learning}), as the Boltzmann learning algorithm also iteratively updates units' states with a probability that depends on an energy function and decreasing temperature~\cite{hinton1986learning, ackley1985learning}, this way of learning, which we described in more detail in Sec.~\ref{sec:qa-for-bm}, is still distinct from the SA algorithm as we use it~\cite{D-WaveSAS} in our QBM (SA), since the former sets the units' values to 1 with a certain probability, regardless of their previous state, while the latter changes their state with a certain probability, even if it was 1 already.}, Deep Boltzmann Machines (DBMs)~\cite{xiaojun2018contractive, yang2022centered} or even convolutional Deep Boltzmann Machines~\cite{salakhutdinov2009deep}. Another way to limit the issues of comparability between CNNs and our QBM model, and possibly also improve its performance by incorporating techniques that make CNNs so successful at processing data with spatial dependencies, such as image classification, involves modifying the model architecture toward a design resembling classical convolutional DBMs, for which training remains demanding~\cite{salakhutdinov2009deep, xiaojun2018contractive, yang2022centered}. This design would consist of multiple layers of hidden units arranged hierarchically and incorporate convolutional connections, both enabling the model to learn increasingly abstract data representations and capture higher-order feature correlations while reducing parameter count~\cite{salakhutdinov2009deep, xiaojun2018contractive, yang2022centered}. This might create a quantum model that can be efficiently trained to perform accurate classification.

\FloatBarrier

\bibliographystyle{ieeetr}
{\small
\bibliography{references}}

\end{document}